\documentclass[aps,prb,twocolumn,groupaddress,floatfix]{revtex4}

\usepackage{graphicx}
\usepackage{dcolumn}
\usepackage{bm}
\usepackage{amsmath}
\usepackage{multirow}

\newcolumntype{f}[1]{D{.}{.}{#1}}

\begin{document}

\title{Membrane Heterogeneity: Manifestation of a Curvature-Induced 
Microemulsion}
\date{\today}
\author{M. Schick}
\affiliation{Department of Physics, University of Washington, Seattle,
  WA 98195}

\begin{abstract} 
To explain the appearance of heterogeneities in the plasma membrane, I 
propose a hypothesis which begins with the observation that fluctuations 
in the membrane curvature are coupled to the difference between 
compositions in one leaf and the other. Because of this coupling, the most 
easily excited fluctuations can occur at non-zero wavenumbers.  When the 
coupling is sufficiently strong, it is well-known that it leads to 
microphase separation and modulated phases. I note that when the coupling 
is less strong, the tendency towards modulation remains manifest in a 
liquid phase that exhibits transient structure of a characteristic size; 
that is, it is a microemulsion.  The characteristic size of the 
fluctuating domains is estimated to be on the order of 100 nm, 
and experiments to verify this hypothesis are proposed.
\end{abstract}

\date{\today}

\maketitle

\section{Introduction}
\label{sec:intro}
Certainly one of the most interesting models of the plasma membrane is 
that, rather than being homogeneous, it is characterized by aggregates of 
saturated lipids and cholesterol which float, like rafts, in a sea of 
unsaturated lipids \cite{simons88}.  An impressive array of experiments 
support this hypothesis \cite{lingwood10}, and limit the size of such 
aggregates in mammalian cells to the order of tens or hundreds of 
nanometers \cite{pralle00,lenne06,schutz00}.
Experiments 
also limit the lifetime of the aggregates so that they are more readily 
described as dynamic domains \cite{pike06}. The hypothesis remains 
controversial, however, due in part to a lack of a firm physical basis for 
the appearance of such domains.

A few explanations have been put forth. One arises from the fact that 
model membrane mixtures of cholesterol and saturated and unsaturated 
lipids readily undergo separation into two liquid phases, one rich in the 
first two components, the other rich in the third \cite{veatch05}. Hence 
rafts might occur in a two-phase region and simply be a domain of one 
phase surrounded by the other. The small size of the domains could then be 
attributed to the effects of the cytoskeleton \cite{yethiraj07}. One 
difficulty with this hypothesis is that it is known that a bilayer of 
composition which mimics the inner leaf of the plasma membrane does not 
undergo phase separation \cite{wang01}, and that the coupling of such a 
leaf to another which does tend to phase separate produces a bilayer in 
which the miscibility transition either occurs at a greatly reduced 
temperature or is eliminated entirely \cite{kiessling06,collins08}.

A second, related, hypothesis is that the inhomogeneities occur in a 
one-phase region, and are simply those fluctuations associated with a 
nearby critical point of two-phase coexistence \cite{honerkampsmith08}. 
Again the size of these fluctuations are proposed to be limited by the 
cytoskelton \cite{matcha11}. This hypothesis is not only subject to the 
criticism that there may be no miscibility phase transition nearby, but 
also to the observation that fluctuations near a critical point exhibit 
little difference between their composition and that of the background 
from which they arise. Consequently they would not easily discriminate 
between different proteins, which is the {\it raison d'{\^e}tre} for the 
raft hypothesis itself.

A third hypothesis is that the fluctuating domains are simply the 
signature of a microemulsion brought about by the presence of a 
line-active agent \cite{brewster09}. The difficulty with this proposal is 
that there is no obvious component to act as such an agent, one which 
would be attracted to the interface between the two phases. In particular 
it is clear that cholesterol is not line active as it prefers the phase 
rich in saturated lipids. Its initial addition to a single liquid phase 
brings about a phase separation \cite{veatch06}, that is, it {\em raises} 
the miscibility transition temperature rather than lowers it as a 
line-active agent would. As cholesterol is not line active, it is posited 
that the unsaturated lipids, which in biological membranes usually have 
one saturated as well as one unsaturated tail \cite{meer08}, can in fact 
play a dual role: as a component of one of the two phases, and as a 
line-active agent between the phases \cite{yamamoto10}. A detailed model 
which encapsulates this idea and explores the effects on this 
microemulsion of the coupling between the two leaves has been explored by 
Hirose et al. \cite{hirose11}. However such a model cannot explain the 
observation of nanoscopic domains in teranary systems that contain no 
lipids with one saturated and one unsaturated tail, as in the system 
dipalmitoy\-lphosphatidylcholine (DPPC), dilauroyl\-phosphatidyl\-choline 
(DLPC), and cholesterol \cite{feigenson01}.

 As this last example illustrates, nanoscopic domains can be brought
about by means other than the action of a line-active agent. Indeed 
microemulsions almost invariably appear in any system which 
manifests modulated phases; they are, in general, the liquid phase to 
which modulated phases melt. It is for this reason that the recent 
observation of modulations of composition in giant unilamellar vesicles 
mimicking 
biological membranes \cite{konyakhina11} is so interesting; it implies 
that such membranes could well display a microemulsion. The questions that 
arise then, concern first, the nature of the interactions within the 
membrane which are responsible for the modulations, and second, the 
characteristic size of the droplets in the two-dimensional microemulsion 
to which the modulated phases melt. A plausible scenario for the 
interactions which give rise to the modulated phases was proposed in the 
seminal work of Leibler \cite{leibler86} and of Leibler and Andelman 
\cite{leibler87}, and I remind the reader in the next section of this 
mechanism which couples the local difference in mole fractions of 
different lipids to the local curvature. There I also estimate the 
characteristic size of the droplets in the microemulsion expected in a 
bilayer with a cytoskeleton, and find it to be on the order of 100 nm. 
Therefore I propose that rafts can be interpreted as the 
characteristic droplets of a curvature-induced microemulsion. Possible 
experiments to verify this hypothesis are proposed.

\section{The Bilayer with coupled curvature-composition fluctuations}

The proposal that fluctuations in curvature and composition are coupled 
goes back to Leibler \cite{leibler86} and  Leibler 
and Andelman \cite{leibler87}. Their work has been extended explicitly to 
bilayers \cite {kodama93,mac94,kumar99}, and I follow the last of these 
here. The basic physical idea is simple. The biological membrane consists 
of a plethora of distinct lipids with different spontaneous curvatures. 
Because of this variation in lipid architecture, fluctuations in the 
difference between local compositions in one leaf and the other couple to 
fluctuations in the curvature of the membrane.  That is, lipids with 
larger headgroups and smaller tails will be attracted preferentially to 
the outer leaf in regions where the membrane bulges outward, while lipids 
with smaller heads and larger tails will be attracted to the inner leaf in 
the same regions. This affinity is directly observed in experiment 
\cite{roux05}.

 For simplicity I consider the cholesterol and saturated lipid as one 
component in a binary system with the unsaturated lipid the other 
component. There are two order parameters representing the differences in 
mol fractions of these two components in the inner leaf, $\Phi_i({\bf 
r}),$ and in the outer leaf, $\Phi_o({\bf r}).$ It is convenient to 
consider the two linear combinations $\phi({\bf r})\equiv (\Phi_i({\bf 
r})-\Phi_o({\bf r}))/2$, and $\psi({\bf r})\equiv (\Phi_i({\bf 
r})+\Phi_o({\bf r}))/2$. The phenomenological free energy consists of 
three pieces. The first is the free energy functional of the planar, 
coupled, bilayer which to second order in the order parameters can be 
written
\begin{equation}
\label{one}
F_{plane}=\int\ d^2r\ \left[\frac{b}{2}(\nabla
\phi)^2+a\phi^2+\frac{b_{\psi}}{2}(\nabla\psi)^2+a_{\psi}\psi^2\right].
\end{equation}
The second
piece is  
the curvature free energy, written here in the Monge representation in terms
of $h({\bf r})$,  the height deviation from the planar configuration,
\begin{equation} 
F_{curv}=\int\ d^2r\ \frac{1}{2}\left[\kappa(\nabla^2h)^2+
\gamma(\nabla h)^2\right],
\end{equation}
with $\kappa$ the bending modulus and $\gamma$ the surface
tension. Lastly there is 
the coupling between the curvature, $\nabla^2h$, and the difference in
compositions between the two leaves, $\phi$ 
\begin{equation}
F_{coupl}=\lambda(b\gamma)^{1/2}\int\ d^2r\ (\nabla^2h)\phi,
\end{equation}
where $\lambda$ is a dimensionless coupling constant. 
In terms of the Fourier transform functions, the total free energy is, up 
to second order,
\begin{eqnarray}
\label{ftot}
F_{tot}&&=\int\
d^2k[(a+\frac{b}{2}k^2)\phi(k)\phi(-k)\nonumber\\
&&+(a_{\psi}+\frac{b_{\psi}}{2}k^2)\psi(k)\psi(-k)\nonumber\\
&&+\frac{1}{2}(\kappa k^4+\gamma k^2)
  h(k)h(-k)\nonumber\\
&&-\lambda(b\gamma)^{1/2} k^2 h(-k)\phi(k)].
\end{eqnarray}
Within mean-field theory, one minimizes the free energy with respect to 
the membrane shape,
$\delta F_{tot}/\delta h(k)=0$,  
%and obtains
%\begin{equation}
%h(k)=\frac{\Lambda}{\gamma}\frac{\phi(k)}{[1+(\kappa k^2/\gamma)]}.
%\end{equation}
and substitutes the resulting height $h[\phi(k)]$ into the free energy, Eq. 
\ref{ftot}, to 
obtain
\begin{eqnarray}
\label{fex}
F_{tot}&=&\int\ d^2k\ \left\{a+\frac{b}{2}
\left( 1-\frac{\lambda^2}{(1+\kappa
  k^2/\gamma)}\right)k^2\right\}\phi(k)\phi(-k)\nonumber\\ 
&+&(a_{\psi}+\frac{b_{\psi}}{2}k^2)\psi(k)\psi(-k).
\end{eqnarray}
This form of the free energy displays everything that is needed. First, the 
wavevector, $k^*$, at
which the composition difference between the two leaves is softest, 
i.e. at which it shows the largest response,  is
the value of $k$ that minimizes
the coefficient of $\phi(k)\phi(-k)$. It is
\begin{eqnarray}
k^*&=&0,\qquad {\rm for}\ \ \lambda<1\nonumber\\ 
\label{kstar}
&=&\left(\frac{\gamma}{\kappa}\right)^{1/2}(\lambda-1)^{1/2},\qquad {\rm for}\ 
\lambda>1.
\end{eqnarray}
Thus the system is softest at a non-zero wavevector when 
$\lambda>1.$ 
This requirement is understood as follows. The 
coupling between the curvature and the difference in compositions 
%is proportional to $-[\lambda(b\gamma)^{1/2}] k^2 h(-k)\phi(k)$ and
%therefore 
favors a soft wavevector $k^*$ that is non-zero. 
As a structure with such a wavenumber 
is curved, and thus of larger area than when flat, this bending 
is opposed by the 
surface tension $\gamma$. Further as regions of different compositions
alternate, their occurrence is opposed by a free energy per unit length
between such 
regions, an energy which is proportional to the coefficient $b$.
Thus the curvature coupling to the composition, measured in terms of the
competing tensions, must be large, {\em i.e.}  $\lambda>1$. 
 The particular consequence of this tendency to display a structure 
characterized by a non-zero wavenumber, $k^*\neq 0$, is determined by the 
coefficient of $\phi(k^*)\phi(-k^*)$ itself, which is equal to
\begin{equation}
a\left(1-\frac{b\gamma}{2\kappa a}(\lambda-1)^2\right)
\end{equation}
When $\lambda$ is not only greater than unity, but is also greater than 
$1+(2a\kappa/b\gamma)^{1/2}$, the coefficient of $\phi(k^*)\phi(-k^*)$ is 
negative, so that the ensemble-average value of $\phi(k^*)$ is non-zero in 
equilibrium; that is the system undergoes microphase separation. The 
resulting phase exhibits either stripes or a triangular array of domains. 
The possible occurrence of these phases was emphasized by the earlier 
works \cite{leibler87,kumar99}, and their possible manifestations in 
coupled bilayers has recently been explored \cite{hirose09}.  Indeed these 
structures have been observed in some simulations of bilayers as predicted 
\cite{stevens05,perlmutter11}. Within mean-field theory, transitions 
between all phases are of first order except at a critical point which can 
occur when the average compositions of the two leaves are identical. 
However even this transition is driven first-order by the large 
fluctuations in the directions of the wavevectors characterizing the 
ordered phases \cite{brazovskii75}, so that all transitions are of first 
order.

The observation that I emphasize here, is that if the system tends toward 
order, but not so strongly as to manifest that order in microphase 
separation, that is if $1+(2a\kappa/b\gamma)^{1/2}> \lambda>1 $, then the 
system will exist in a fluid phase, but a fluid that still reflects a 
tendency towards order. That tendency is manifest in its composition 
fluctuations, which are strongest at a non-zero wavevector. Its structure 
is reflected in the structure factor, $S(k)$ where $S(k)^{-1}$ is the 
coefficient of $\phi(k)\phi(-k)$. The structure factor has a peak at $k^*$ 
which is non-zero when $\lambda-1>0.$ The structure is also reflected in 
the correlation function, $g({\bf r}),$ which is the inverse Fourier 
transform of $S({\bf k})$. In the interesting regime in which 
$(2a\kappa/b\gamma)^{1/2}> \lambda-1>0 $, and for small wavenumbers 
$\kappa k^2/\gamma<1$ it is straightforward to show that $g(r)$ behaves, 
for large $r$ like
%\begin{eqnarray}
%g(r)&=&\frac{1}{(2\pi)^2}\int_0^{2\pi}d\theta\int_0^{\infty}dkk
%\frac{e^{ikr\cos\theta}}{a+c_2k^2+c_4k^4}\nonumber\\
%&=&\frac{1}{2\pi c_4[k_1^2-(k_1^2)^*]}\int_0^{\infty}\left[\frac{J_0(kr)}{k^2-k_1^2}-
%\frac{J_0(kr)}{k^2-(k_1^2)^*}\right]kdk\nonumber\\
%&=&\frac{1}{4(ac)^{1/2}(1-c_2^2/4ac_4)^{1/2}}Re\ H_0^{(1)}(ik_1r)
%\end{eqnarray}
%where $H_0^{(1)}$ is the Hankel function of the first kind and
%$k_1\equiv k_c+i\xi^{-1}$ with $k_c$ and $\xi$ given explicitly below.
%This correlation function behaves for large $r$ like 
$g({\bf r})\approx r^{-1/2}\exp(-r/\xi)\sin(k_cr+\delta),$ with $k_c$ and 
$\xi$ given explicitly below and $\delta$ a phase of no interest. The 
exponential damping, with a characteristic correlation length $\xi$, 
implies that the system is disordered, a liquid.  The oscillatory function 
introduces an {\em additional} length, $k_c^{-1}$, and this shows that the 
liquid is structured. It is this property of a liquid, to display structure 
at a length scale in addition to that of the correlation length, that is 
characteristic of a microemulsion. The correlation length, $\xi$, and 
characteristic wavenumber, $k_c$, are given by
\begin{eqnarray} 
2\xi^{-2}
&=&\left[\left(\frac{2a\kappa}{\lambda^2b\gamma}\right)^{1/2} 
-\frac{1}{2}\left(1-\frac{1}{\lambda^2}\right)\right] 
\left(\frac{\gamma}{\kappa}\right),\\ 
\label{kc}
2k_c^2
&=&\left[\left(\frac{2a\kappa}{\lambda^2b\gamma}\right)^{1/2} 
+\frac{1}{2}\left(1-\frac{1}{\lambda^2}\right)\right] 
\left(\frac{\gamma}{\kappa}\right). \end{eqnarray}
The correlation length $\xi$ is equal to the characteristic distance 
$k_c^{-1}$ at the Lifshitz line at which $\lambda=1$. Consequently the 
microemulsion structure is strongly damped. However the correlation length 
is larger than the characteristic distance for $\lambda>1$ which indicates 
that characteristic oscillations in the fluid are manifest before being 
damped out.
%In this regime $k_c$, $\xi^{-1},$ and $k^*$,
%the wavenumber at which the structure factor peaks, are related by
%\begin{equation}
%k_c^2-\xi^{-2}=(k^*)^2=\left(\frac{\gamma}{2\kappa}\right)\left(1-\frac{1}{\lambda^2}\right).
%\end{equation}
From Eq. (\ref{kc}) we see that the characteristic distance, $k_c^{-1},$ is on the 
order of, or larger, than $(2\kappa/\gamma)^{1/2}$.  As typical values of 
the bending modulus and the tension of a membrane in the presence of a 
cytoskeleton are\cite{dai99} $\kappa\approx 2.7\times 10^{-19}$Nm 
and 
 $\gamma\approx 
2\times 10^{-5}$N/m, the characteristic size of the fluctuating regions is 
on the order, or greater than, $10^{-7}$m, or 100 nm. This mean-field 
estimate indicates that the proposed mechanism could account for regions 
of the observed size.

\section{Discussion}
%\subsection{Formalism}
%\label{sec:meth}
I have proposed that inhomogeneities in the plasma membrane, and those 
observed in model membranes, are microemulsions brought about by the 
coupling of curvature to the difference in composition of the two leaves. 
This hypothesis of a microemulsion avoids the difficulties 
associated with ascribing such inhomogeneities either to phase separation 
or to the fluctuations associated with a critical point. As noted earlier, 
that a biological membrane could display a microemulsion is strongly 
indicated by the recent observation in giant unilamellar vesicles of 
modulations of composition in a four-component system consisting of 
distearoyl\-phosphatidyl\-choline (DSPC), dioleoyl\-phosphatidyl\-choline 
(DOPC), 1-palmitoyl\- 2-oleoyl\-phosphatidyl\-choline (POPC), and 
cholesterol \cite{konyakhina11}.
%As is
%well known in the literature of oil, water, and amphiphile systems,
%modulated phases are invariably accompanied by microemulsions in a
%nearby part of the phase diagram \cite{gompper94}. Indeed microemulsions
%are the liquid phase to which modulated phases become unstable.
%Previously the occurrence of such a
%microemulsion was attributed to a line-active agent, a hybrid lipid
%\cite{yamamoto10}, but as noted above, this cannot
%explain the observation of inhomogeneities in systems without
%such lipids\cite{feigenson01}.
That curvature is strongly indicated as the mechanism responsible for 
bringing about the modulations is also evidenced by other results of 
this four-component system. With fixed mole fractions of DSPC and cholesterol, 
the relative mole fraction $\rho\equiv $ [DOPC]/([DOPC]+[POPC]) was 
varied. One expects that the difference in spontaneous curvature between 
DSPC and DOPC is greater than that between DSPC and POPC. Therefore 
increasing the fraction $\rho$ from small values should drive the system 
towards a modulated phase, and this is indeed what is observed. Similarly 
decreasing the value of $\rho$ from the modulated phase is expected to 
cause it to become unstable to a fluid phase, one which would appear 
uniform to fluorescence microscopy. Again this is what is observed. 

Additional support for the mechanism proposed here would be provided 
by an estimate of the value of  $\lambda$ which
characterizes the strength of the coupling $\lambda(b\gamma)^{1/2}$ between
the curvature and the difference of lipid mol fractions. It must be  
on the order of unity  for a microemulsion to occur.   
Leibler and Andelman \cite{leibler87} in their original paper, and later
Liu et al. \cite{liu05}, reasonably assume that the energy per unit
length,  $\lambda(b\gamma)^{1/2}$, should be set equal to $\kappa\delta
H$ where $\delta H$ is the difference in spontaneous curvatures of
cholesterol-rich raft domains and the phospholipid background. With the
coupling $b$ of Eq. (\ref{one}) of order $k_BT$, this yields a simple
expression for the dimensionless coupling
$\lambda=[\kappa/k_BT][k_BT(\delta H)^2/\gamma]^{1/2}$. Liu et
al. estimated $\delta H$ to be on the order of $10^6$m$^{-1}$ and took 
$\gamma=3.1\times 10^{-6}$N/m and $\kappa=400k_BT$. These values yield
an estimate for $\lambda$ of about 14 which would imply that such
membranes should always display a modulated phase, contrary to
experiment. This negative result would support their conclusion, 
which they reached by a
slightly different argument, that the coupling
between curvature and concentration fluctuations could not explain raft
phenomena. However, if one utilizes the same difference in spontaneous
curvature, the larger surface tension $\gamma =2\times 10^{-5}$N/m of
Dai and Sheetz \cite{dai99} and the same reference's smaller 
bending modulus
$\kappa=2.7\times 10^{-19}$Nm=\ 66$k_BT$, then one obtains the estimate 
$\lambda=0.94.$ This shows that it is certainly plausible that the
coupling has the correct order of magnitude to bring about the existence of a 
microemulsion.

The coupling that is assumed in this paper to produce the microemulsion, 
one between curvature and the {\em difference} in the compositions between 
the two leaves, predicts that saturated lipid-rich and -poor 
regions in the two leaves are {\em anticorrelated}. This is in line with 
the results of coarse-grained simulations of the ternary mixture of 
diarachidoyl\-phosphatidyl\-choline, dilinoleoyl\-phosphatidyl\-choline, 
and cholesterol \cite{perlmutter11} of which the first two components 
differ markedly in curvature. The simulations show a modulated stripe 
phase in which the stripes in the two leaves are indeed anticorrelated. 
This anticorrelation has interesting consequences for the microemulsion. 
For example, an area in the outer leaf rich in saturated lipids and 
cholesterol would, due to the damped oscillations in composition, be 
bordered by a region rich in unsaturated lipids. These areas are 
anticorrelated with regions of the inner leaf; i.e. the above areas in the 
outer leaf would face in the inner leaf a region rich in unsaturated 
lipids bordered by one rich in saturated lipids and cholesterol. 
It would be most interesting, of course, to determine whether domains, either 
in the microemulsion or modulated phases, are indeed anticorrelated. 
Perhaps this could be accomplished in the modulated phases by tagging the 
lipids on the inner and outer leaves of the vesicles with different
dyes. Another possibility would be to observe images obtained from the system 
after being subjected to freeze
fracture, as in ordinary microemulsions \cite{jahn88}.

The hypothesis makes additional predictions. For example the 
ternary system POPC, DSPC, and cholesterol does not undergo macroscopic 
phase separation but does show nanodomains \cite{heberle10}. One also 
knows that the ternary system POPC, DPPC, and cholesterol does not undergo 
macroscopic phase separation \cite{veatch03}. As the difference in 
spontaneous curvature of the two cholines in the latter is certainly 
larger than in the former, one would expect nanodomains to be present,
and this can be ascertained by F{\"o}rster resonance energy transfer
(FRET). It 
would also be of interest to consider the ternary systems in which DPPC is 
replaced by dimyristoyl\-phosphatidyl\-choline, or 
dilauroyl\-phosphatidyl\-choline as these would also increase the 
curvature difference. The dependence of the characteristic wavenumber 
of the domains, 
 $k_c$ Eq. (\ref{kc}), on surface tension suggests that 
the domain size, obtained experimentally by FRET, 
could be varied by controlling the tension in experiment \cite{portet12}. 

%\begin{figure}[t]
%\includegraphics[scale=0.35,clip]{Figs/Ge_DQtQ0.eps}
%\caption{\label{fig:ge_corr}
%This is the caption
%}
%\end{figure}
\begin{acknowledgments}
This work was supported by the National Science Foundation under grant 
DMR-0803956.  I readily acknowledge stimulating conversations with present 
and former members of the Keller Group at the University of Washington, 
and helpful communications with S. Komura, D. Andelman, M. H{\"o}mberg, 
M. M{\"u}ller, and G. Feigenson.
\end{acknowledgments}
%\bibliography{lipids7}
%\bibliographystyle{apsrev}

%\end{thebibliography}

\end{document}